\documentclass[sigconf]{acmart}

\newtheorem{mydef}{Definition}

\usepackage{multirow}
\usepackage{yfonts}
\usepackage{longtable}
\usepackage{arydshln}
\usepackage{makecell}
\usepackage{enumerate}
\usepackage{amsmath}
\usepackage{enumitem}
\usepackage{natbib}
\usepackage{graphicx}  
\usepackage{bm}

\usepackage[ampersand]{easylist}

\usepackage{subcaption}

\AtBeginDocument{%
  \providecommand\BibTeX{{%
    \normalfont B\kern-0.5em{\scshape i\kern-0.25em b}\kern-0.8em\TeX}}}

\copyrightyear{2022} 
\acmYear{2022} 
\setcopyright{acmlicensed}\acmConference[CIKM '22]{Proceedings of the 31st ACM International Conference on Information and Knowledge Management}{October 17--21, 2022}{Atlanta, GA, USA}
\acmBooktitle{Proceedings of the 31st ACM International Conference on Information and Knowledge Management (CIKM '22), October 17--21, 2022, Atlanta, GA, USA}
\acmPrice{15.00}
\acmDOI{10.1145/3511808.3557300}
\acmISBN{978-1-4503-9236-5/22/10}

\DeclareMathOperator*{\argmin}{arg\,min}

\begin{document}

\title{Dynamic Causal Collaborative Filtering}

\author{Shuyuan Xu}
 \affiliation{
  \institution{Rutgers University}
  \institution{New Brunswick, NJ, US}
    }
 \email{shuyuan.xu@rutgers.edu}

\author{Juntao Tan}
 \affiliation{
  \institution{Rutgers University}
  \institution{New Brunswick, NJ, US}
    }
 \email{juntao.tan@rutgers.edu}
 
 \author{Zuohui Fu}
 \affiliation{
  \institution{Rutgers University}
  \institution{New Brunswick, NJ, US}
    }
 \email{zuohui.fu@rutgers.edu}
 
 \author{Jianchao Ji}
 \affiliation{
  \institution{Rutgers University}
  \institution{New Brunswick, NJ, US}
    }
 \email{jianchao.ji@rutgers.edu}
 
 \author{Shelby	Heinecke}
 \affiliation{
  \institution{Salesforce Research}
  \institution{Palo Alto, CA, US}
    }
 \email{shelby.heinecke@salesforce.com}

\author{Yongfeng Zhang}
 \affiliation{
  \institution{Rutgers University}
  \institution{New Brunswick, NJ, US}
    }
 \email{yongfeng.zhang@rutgers.edu}

\renewcommand{\shortauthors}{Shuyuan Xu et al.}

\begin{abstract}


Causal graph, as an effective and powerful tool for causal modeling, is usually assumed as a Directed Acyclic Graph (DAG). However, recommender systems usually involve feedback loops, defined as the cyclic process of recommending items, incorporating user feedback in model updates, and repeating the procedure. As a result, it is important to incorporate loops into the causal graphs to accurately model the dynamic and iterative data generation process for recommender systems. However, feedback loops are not always beneficial since over time they may encourage more and more narrowed content exposure, which if left unattended, may results in echo chambers. As a result, it is important to understand when the recommendations will lead to echo chambers and how to mitigate echo chambers without hurting the recommendation performance.

In this paper, we design a causal graph with loops to describe the dynamic process of recommendation. We then take Markov process to analyze the mathematical properties of echo chamber such as the conditions that lead to echo chambers. Inspired by the theoretical analysis, we propose a Dynamic Causal Collaborative Filtering ($\partial$CCF) model, which estimates users' post-intervention preference on items based on back-door adjustment and mitigates echo chamber with counterfactual reasoning. Multiple experiments are conducted on real-world datasets and results show that our framework can mitigate echo chambers better than other state-of-the-art frameworks while achieving comparable recommendation performance with the base recommendation models.
\end{abstract}



\maketitle

\section{Introduction}\label{sec:intro}

Recommender Systems (RS) aim to provide personalized services for users, occupying an expanding role in a wide range of applications, such as e-commerce, video streaming, social media and online job markets. Recent efforts have been made towards causal recommendation, which incorporates causation into recommender systems to enable causal inference over some critical aspects, such as eliminating bias \cite{bonner2018causal,schnabel2016recommendations}, promoting fairness \cite{li2021towards, zhang2018fairness,ge2022explainable}, improving robustness \cite{li2022causal} and enhancing explainability \cite{tan2021counterfactual,ghazimatin2020prince,tran2021counterfactual}.

Causal graph is an effective and powerful tool that enables researchers to estimate desired values \cite{xu2021causal,glymour2016causal}. Typically, the causal graph is constructed as a directed acyclic graph (DAG) depicting the data generation process. However, in real-word recommender systems, 
the data generation process usually spans over a period of time. Within this period, the recommendation results made by the system can have a great impact over users' interests and decision preference, and in turn influence the feedback that the system receives. This dynamic process is called a \emph{feedback loop} \cite{jiang2019degenerate, ge2020understanding}, which may not be accurately captured by a DAG causal graph.
Therefore, introducing loops into the causal graph design may allow causal models to understand the dynamic data generation process in recommender systems more comprehensively.

While the causal graph with loops can better capture the dynamic and iterative data generation process in recommender systems, the presence of feedback loops is not always beneficial. Specifically, feedback loops may narrow the user's interest towards certain contents, which may further lead to decreased engagement with the system \cite{chaney2018algorithmic}. Additionally, feedback loops, if left unattended, may also result in what is known as \emph{echo chambers} \cite{ge2020understanding,chaney2018algorithmic}.
In general, echo chambers describe the homogenization of social communities \cite{sunstein2009going} that occur as a result of feedback loops. This homogenization isolates users in information echo chambers, severely limiting their information exposure. The existence of echo chambers has been validated on recommender systems with respect to e-commerce \cite{ge2020understanding}, online media \cite{nguyen2014exploring,allen2017effects,hilbert2018communicating}, and social networking \cite{risius2019towards,stoica2019hegemony}. 

Take movie recommendation as an example to better understand the consequences of echo chambers.
Suppose a user has provided positive feedback towards action movies, and the system is able to learn that the user would be highly possible to be interested in action movies. If the system does not properly handle echo chamber when making recommendations, it may repeatedly recommend other action movies and will likely receive positive feedback in the short term. However, over time the system will gradually converge to only showing action movies and make the user dissatisfied eventually. Given the ubiquity of recommender systems in our daily-life, it is important to understand when a recommender system results in echo chambers and how to mitigate echo chambers while maintaining the recommendation performance.

In this paper, we particularly seek to answer the following three questions regarding the task of mitigating echo chambers: 1) how to design a causal graph with loops to describe the dynamic and iterative data generation process of recommendation, 2) if a recommender system does not take care of echo chamber, when will the system result in echo chamber, and 3) how to mitigate echo chambers without hurting the recommendation performance.

Specifically, we first design a causal graph as a directed graph with loops to describe how the data is generated in a dynamic and iterative manner. Based on our constructed causal graph, we conduct mathematical analysis to understand the conditions that a system will result in echo chamber. Concretely, we represent the dynamic recommendation process as a Markov Process. Then we categorize user behavior into three types and check the existence of homogeneity brought by echo chambers. The analysis confirms the soundness of our proposed causal graph (Section \ref{sec:echochambers}).

In addition to the above theoretical contributions, our work also provides essential technical contributions. We propose a Dynamic Causal Collaborative Filtering ($\partial$CCF) framework for recommendation to mitigate echo chambers.
More specifically, we apply the back-door adjustment to estimate the post-intervention effect based on the unfolded causal graph. Inspired by our theoretical analysis, we apply counterfactual reasoning to mitigate the echo chamber effect while retaining the recommendation performance (Section \ref{sec:model}).
We conduct experiments over two real-world datasets to evaluate the effect of echo chambers by measuring the change of content diversity \cite{ge2020understanding}. The results confirm that our framework is capable of mitigating echo chambers while obtaining comparable recommendation performance with the base recommendation models.

The key contributions of our paper can be summarized as follows:
\begin{itemize}[leftmargin=*]
    \item We design a causal graph with loops to represent the dynamic data generation process of recommender systems.
    \item We represent the user-system interaction as a Markov Process to understand the conditions that lead a system to echo chambers if the system does not properly handle echo chamber.
    \item We propose Dynamic Causal Collaborative Filtering ($\partial$CCF), which takes the back-door adjustment to estimate user preferences and applies counterfactual reasoning to mitigate echo chambers.
    \item Experiments on two real-world datasets show that our framework can mitigate echo chambers better than other methods while maintaining comparable recommendation performance with the base recommendation models.
\end{itemize}

The remainder of this paper is organized as follows. We discuss the related work in Section \ref{sec:relatedworks}. In Section \ref{sec:pre}, we introduce our designed causal graph and the theoretical analysis of echo chambers. We show the details of our proposed dynamic causal collaborative filtering framework in Section \ref{sec:model}. We describe our experiments on real-world datasets and discuss the results in Section \ref{sec:experiments}. Finally, in Section \ref{sec:conclusion}, we conclude the work and discuss future directions.

\section{Related Work}\label{sec:relatedworks}

\subsection{Causal Recommendation}
Causal machine learning have been explored to tackle some critical problems in recommender systems. For example, researchers have leveraged causal models to enhance explainability \cite{tan2021counterfactual, ghazimatin2020prince, tran2021counterfactual, tan2022learning,zhang2020explainable}, promote fairness \cite{li2021towards, zhang2018fairness,ge2022explainable,li2022fairness}, eliminate bias \cite{bonner2018causal, liu2020general, schnabel2016recommendations, yang2018unbiased, zheng2021disentangling, qin2020attribute, wang2020causal, xu2021deconfounded}, improve robustness \cite{li2022causal,ge2022survey}, estimate the recommendation uplifts \cite{sato2019uplift, sato2019action, sato2020unbiased}, and enable counterfactual reasoning for data augmentation \cite{xiong2021counterfactual, wang2021counterfactual, yang2021top, zhang2021causerec}. 
Across the various causal recommendation models, causal graph is shown to be a powerful tool that enables counterfactual reasoning. Existing models usually assume that causal graphs are directed acyclic graphs (DAG).
However, the data used to train recommender systems is generated in a continuous and dynamic manner, which cannot be captured by a DAG as a single snapshot. Therefore, a causal graph with loops is needed to describe the dynamic recommendation scenario.

\subsection{Echo Chambers}
Echo chambers have been studied under several contexts such as social networks \cite{bakshy2015exposure, chitra2020analyzing, garimella2018political,risius2019towards,stoica2019hegemony} and opinion dynamics \cite{banisch2019opinion,perra2019modelling,krueger2017conformity}. Meanwhile, the importance of content diversity has been explored to address user polarization problems \cite{aslay2018maximizing, kaminskas2016diversity, kunaver2017diversity,matakos2020tell,nguyen2014exploring}. In recent years, the study of feedback loops and echo chambers in recommender systems has received a lot of attention \cite{moller2018not,ge2020understanding,chaney2018algorithmic,jiang2019degenerate}. For example, \citet{ge2020understanding} analyzed the property of echo chambers in e-commerce recommender systems.
Existing works have demonstrated preliminary progress toward addressing feedback loops and mitigating echo chambers. For example, \citet{jiang2019degenerate} use the dynamic system framework to model user interest and treat interest extremes as a degeneracy point of the system. However, they assume that users and items are independent of each other, which is inconsistent with the principle of how collaborative filtering works. \citet{kalimeris2021preference} study the similar phenomenon and solve it by learning a stable fixed point, where user preferences do not change in response to the system recommendations. However, this work is limited to matrix factorization-based recommendation and does not consider causality. Instead, our work aims to mitigate echo chambers in a more general setting based on causal reasoning.

\section{Theoretical Analysis}\label{sec:pre}
In this section, we introduce our designed causal graph and the theoretical analysis of under what conditions will recommender systems result in echo chamber.

\subsection{The Causal Graph}\label{sec:causalgraph}

\begin{figure}
    \centering
    \includegraphics[scale=0.9]{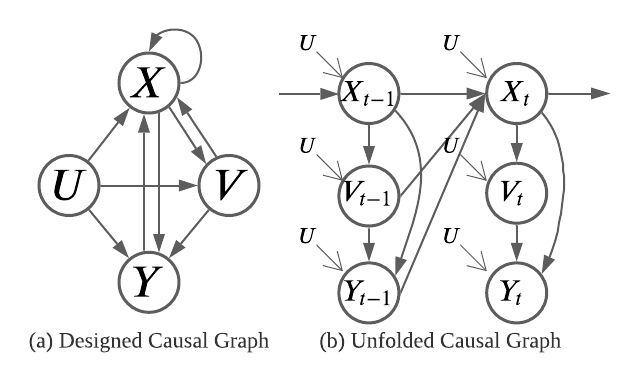}
    \vspace{-20pt}
    \caption{In the causal graph, $U$ is user, $V$ is the exposed item, $X$ is user interaction history, and $Y$ is preference score. (a) is our designed causal graph, more details are introduced in Section \ref{sec:causalgraph}. (b) is the temporal unfolded causal graph, which guides the design of the proposed framework. We reorganize causal graph using $U$ as exogenous variable to make it clearer. More details are introduced in Section \ref{sec:model}.}
    \vspace{-10pt}
    \label{fig:causalgraph}
\end{figure}

Figure \ref{fig:causalgraph}(a) shows our designed causal graph for dynamic recommendation. Existing works usually assume the causal graph as a directed acyclic graph (DAG). However, as we mentioned before, a causal graph with loops may better capture the mechanism in recommender systems. 
We use uppercase letters (e.g., $U$) to denote random variables and lowercase letters (e.g., $u$) to denote the corresponding specific values. We explain the rationality of our designed causal graph from the view of data generation as follows:
\begin{itemize}[leftmargin=*]
    \item Node $U$ represents the user variable. More specifically, we take user ID as the variable values.
    \item Node $V$ represents the exposed item variable. Similarly, we take item ID as the variable values.
    \item Node $Y$ represents preference score. In this work, for simplicity, we consider variable $Y$ as a binary variable, i.e., $Y=1$ for \textit{like} and $Y=0$ for \textit{dislike}.
    \item Node $X$ represents the user interaction history. More specifically, it is a sequence of item IDs and the corresponding preference scores from the user.
    \item Edges $\{U, X\}\rightarrow V$ denote that the exposed item $V$ is determined by user $U$ and the user interaction history $X$.
    \item Edges $\{U,X,V,Y\}\rightarrow X$ denote the dynamic generation process of the user interaction history $X$.
    When an exposed item $V=v$ is generated, user's preference towards $v$ as $Y=y_v$ combined with the exposed item $v$ will be integrated into the history and update the previous history as a new history $X$, which is used to estimate the next exposed item. These edges representing the iterative and dynamic process construct the feedback loop in our designed causal graph.
    \item Edges $\{U,X,V\}\rightarrow Y$ denote that the user preference score is determined by user $U$, user interaction history $X$ and the exposed item $V$. The preference of a user-item pair is not only determined by the corresponding $(u,v)$ pair but also affected by the previous interaction history. Recall the example in Section \ref{sec:intro}, for a user who is interested in action movies, it is highly likely to obtain a positive feedback for an action movie in the beginning, but if the user has already watched so many action movies, the user may be tired of action movies and provide a negative feedback towards a new one.
\end{itemize}
From the causal graph, we can see four loops $X\rightarrow X$, $X\rightarrow V \rightarrow X$, $X\rightarrow Y \rightarrow X$ and $X\rightarrow V \rightarrow Y \rightarrow X$. These loops represent the dynamic and iterative updating process of the user interaction history, embodying the feedback loop in recommender systems.

\subsection{Understanding Echo Chambers}\label{sec:echochambers}
In this section, we will provide a theoretical analysis of echo chambers in recommender systems to understand under what condition the system will lead to echo chambers if the system does not properly handle echo chambers. The following analysis targets recommendation models that do not consider echo chambers.

\subsubsection{\textbf{Problem Setting and Notations}} Generally speaking, recommender systems, especially Collaborative Filtering (CF) based models, explicitly or implicitly learn the similarity between items for recommendation \cite{ekstrand2011collaborative,zhang2019deep,zhang2020explainable}. For example, 
Matrix Factorization (MF) models such as \cite{koren2009matrix} predict $\mathbf{u}^\intercal\mathbf{v}$ for recommendation, where $\mathbf{u}$ and $\mathbf{v}$ are the user and item latent factors, and such models learn the similarity between items that are projected by user latent factor $\mathbf{u}$ (i.e., for a certain user $u$, if $\mathbf{u}^\intercal\mathbf{v_1}$ is similar to $\mathbf{u}^\intercal\mathbf{v_2}$, then $v_1$ is similar to $v_2$ in terms of user $u$).
Since the recommendation models are able to measure the similarity, models are capable of grouping items based on the corresponding similarity measurement. Suppose there are $n$ items and $d$ groups ($n\gg d$), we denote the set of items as $\mathcal{I}=\{i_1, i_2, \cdots, i_n\}$ and the set of groups as $\mathcal{C}=\{c_1, c_2, \cdots, c_d\}$. We denote the grouping function as $A: \mathcal{I}\rightarrow \mathcal{C}$, which returns a group for each item based on similarity.

If the system does not consider echo chambers carefully, then similarity-driven recommendation tends to recommend items that are similar to previously interacted items, i.e., they may belong to the same group as the items in the user's interaction history \cite{moller2018not,kaminskas2016diversity,kunaver2017diversity}.
This tendency stems from the principle of CF-based models that users have similar tastes in the past are likely to have similar interests in the future \cite{ekstrand2011collaborative}.
For example, a simple MF model such as \cite{koren2009matrix} recommends items with higher preference which is calculated as $\mathbf{u}^\intercal\mathbf{v}$. A well-trained MF model produces high preference scores on items within the interaction history, thus the recommended items with high preference scores would be similar to one or more items within the interaction history since they are both similar to the user embedding $\mathbf{u}$.
Since the grouping is based on the similarity learned by the model, a recommended item will belong to the same group with its similar items in the interaction history.

We denote the capacity of items in the interaction history as $m$ (echo chambers occur in the repetitive behaviors \cite{ge2020understanding} thus the cold-start scenarios are not within the scope of our discussion). It is worth clarifying that $m$ can be infinity, but in practice, the items interacted a long time ago 
may have little influence on current predictions. 
We represent the history as a sequence of items $\{h_1, h_2, \cdots, h_m\},$ where $h_i\in \mathcal{I}$ (if the number of valid items $\tilde{m}$ in the history is less than $m$, then the first $m-\tilde{m}$ items would be empty). When a new item is interacted and added to the user history, the first item would be removed from the history to keep the history at length $m$. Applying the grouping function $A$ to the item sequence, we obtain a historical group sequence $\{A(h_1), A(h_2), \cdots, A(h_m)\}, A(h_i)\in \mathcal{C}$, which can be simplified as $\{A_1, A_2, \cdots, A_m\}, A_i\in \mathcal{C}$. 
Since there are $d$ groups in total, there are $d^m$ possible historical group sequences. When the user interacts with a new item, both the item sequence and the historical group sequence will be updated. 
Since the recommendation models that do not consider echo chambers tend to recommend items that are similar with historical items (i.e., most likely belong to the same group), the group of the new item only depends on the current historical group sequence. 
Therefore, we can formulate this iterative process as a Markov chain. 

We consider the states in the Markov chain as historical group sequences, therefore, there are $d^m$ different states. 
The transition probabilities are determined by the policy of interacting with new items. 
To analyze the occurrence of echo chambers for different users, we categorize user behavior into the following three types.
\begin{enumerate}
    \item Users only interact with recommended items.
    \item Users completely ignore the recommendations.
    \item Users interact with both recommended and not recommended items.
\end{enumerate}
We will separately analyze the above three categories from the perspective of echo chambers.

\subsubsection{\textbf{Type 1}}
When users only interact with recommended items, the transition probability is fully determined by the recommendation model. 
As we mentioned before, recommendation models that do not consider echo chambers tend to recommend items belonging to the same group as historical items. We can define a general one-step transition probability as follows:
{\small
\begin{equation}\label{eq:transition1}
\begin{split}
    &P(\{A'_1, \cdots, A'_m\}|\{A_1, \cdots, A_m\}) \\
    =& \begin{cases}
    [A'_m \in \{A_1, \cdots, A_m\}]P_{A'_m} &\text{if $A'_j=A_{j+1},j=1,\cdots, m-1$}\\
    0 &\text{otherwise}
    \end{cases}
\end{split}
\end{equation}}
where $[P]$ represents the Iverson bracket, which maps a statement into a binary value (i.e., takes value $1$ if the statement $P$ is True and 0 otherwise) and $P_{A'_m}$ is the probability that the recommended items belong to group $A'_m$. Based on the one-step transition probability defined as Eq.\eqref{eq:transition1}, we are able to discover some special states. If $A_1=A_2=\cdots=A_m$, we have
\begin{equation}\label{eq:absorbing}
\begin{split}
    &P(\{A'_1, \cdots, A'_m\}|\{A_1, \cdots, A_m\}) \\
    =& \begin{cases}
    1 & \text{if $A'_1=A'_2=\cdots=A'_m=A_i$}\\
    0 & \text{otherwise}.
    \end{cases}
\end{split}
\end{equation}
We call these special states \textit{absorbing states}; once users enter these states, they can no longer exit. 
\begin{mydef}\cite{tolver2016introduction} \label{def:absorbing}
(Absorbing Markov Chains) A Markov chain is an absorbing Markov chain if:
\begin{enumerate}[label = {(\alph*)}]
    \item there is at least one absorbing state and
    \item it is possible to go from any state to at least one absorbing state in a finite number of steps.
\end{enumerate}
\end{mydef}

Based on the transition probabilities shown in Eq.\eqref{eq:absorbing}, there are $d$ absorbing states in total. Meanwhile, based on Eq.\eqref{eq:transition1}, we can derive that $P(\{A'_1, \cdots, A'_m\}|\{A_1, \cdots, A_m\})>0$ if $A'_m=A_m, A'_j=A_{j+1},j=1,\cdots, m-1$. 
Therefore, it is possible to go from any states to an absorbing state in $m$ steps. According to Definition \ref{def:absorbing}, we can claim that in this setting, we have an absorbing Markov chain.

As an absorbing Markov chain with $d$ absorbing states, the transition matrix $P$ can be transformed into a block matrix, producing the following $k$-step transition matrix when $k\to\infty$ \cite{tolver2016introduction}:
\begin{equation}
    P=
    \begin{bmatrix}
    Q & R \\
    \mathbf{0} & I_d
    \end{bmatrix} \Rightarrow
    \lim_{k \to \infty} P^k=
    \begin{bmatrix}
    \mathbf{0} & (I-Q)^{-1}R\\
    \mathbf{0} & I_d
    \end{bmatrix}
\end{equation}

In summary, if a user keeps interacting with the recommended items, the interacted items will eventually belong to the same group and will not include other groups any more. This is how echo chambers lead to homogeneity and polarization.

\subsubsection{\textbf{Type 2}}
If users completely ignore the recommendation model, then each group has a nonzero probability to be interacted. We can define the one-step transition probability as follows:
\begin{equation}\label{eq:transition2}
\begin{split}
    &P(\{A'_1, \cdots, A'_m\}|\{A_1, \cdots, A_m\}) \\
    =& \begin{cases}
    P'_{A'_m} &\text{if $A'_j=A_{j+1},j=1,\cdots, m-1$}\\
    0 &\text{else}
    \end{cases}
\end{split}
\end{equation}
where $P'_{A'_m}$ is the probability that user chooses to interact with the item in the $A'_m$ group without the influence of recommendation.

Based on the one-step transition probability shown in Eq.\eqref{eq:transition2}, we can calculate the $k$-step transition matrix. 
Since $P'_{A'_m}$ is a nonzero value for any group, there is no absorbing state, which means that the interaction history will not fall into the same group and remain unchanged. 
We take uniform distribution as a simple example (i.e., $P'_{A'_m}=1/d$), in this case, we can obtain the following $k$-step transition matrix if $k$ is large enough:
\begin{equation}
    P^k = \Big[\frac{1}{d^m}\Big]_{d^m\times d^m},~~\text{if $k\geq \log m + 1$}
\end{equation}

For users in this type, users will avoid the drawbacks brought by echo chambers as long as each group has a nonzero probability to be interacted. 
However, considering the extremely large scale of items in practice, completely ignoring the recommendations will lose the benefits offered by personalized recommender systems.

\subsubsection{\textbf{Type 3}}
For the last type of users, we denote probability $p\in (0,1)$ as following recommendation models and probability $1-p$ as not following recommendation models. 
We can then define the one-step transition probability as follows:
{
\begin{equation}\label{eq:transition3}
\begin{split}
    &P(\{A'_1, \cdots, A'_m\}|\{A_1, \cdots, A_m\}) \\
    =& \begin{cases}
    P^f_{A'_m}p+P'_{A'_m}(1-p) &\text{if $A'_j=A_{j+1},j=1,\cdots, m-1$}\\
    0 &\text{else}
    \end{cases}
\end{split}
\end{equation}}%
where $P^f_{A'_m}=[A'_m \in \{A_1, \cdots, A_m\}]P_{A'_m}$ represents the probability of the recommended items belonging to group $A'_m$ as Eq.\eqref{eq:transition1}, and $P'_{A'_m}$ is the probability that user chooses to interact with the item in the $A'_m$ group without the influence of recommendation.

Notice that the Markov chain is irreducible. In other words, it is possible to result in any state from an arbitrary starting state after $m$ steps, because $P(\{A'_1, \cdots, A'_m\}|\{A_1, \cdots, A_m\})>0$, $\forall A'_m\in \mathcal{C}, A'_j=A_{j+1},j=1,\cdots, m-1$. 
Since every state is accessible from any other state, In Type 3, the effect of echo chambers could be alleviated to some extent. 
When $p$ is close to 1, the effect of echo chamber would be increased. 
When $p$ is close to 0, the effect of echo chamber would be mitigated, but it also loses the benefits of personalized recommendation. 
In other words, if probability $p$ is properly chosen, users will not only benefit from the recommendation results but also avoid homogeneity brought by echo chambers.

Given the evidence of echo chambers in some existing works \cite{ge2020understanding,chaney2018algorithmic}, the above analysis based on our designed causal graph confirms the rationality of the causal graph. 
Based on our analysis, if the recommender system does not consider echo chambers, whether recommendation will lead to echo chamber is determined by user behavior. 
If the user actively explores the items of interest instead of only passively interacting with recommended items, the user may not be affected by echo chambers. If the user only passively interacting with the recommended items and the recommendation models do not consider echo chambers, then the user is likely to be affected by echo chambers.

In real-world applications, we cannot control the user behaviors. Therefore, mitigating echo chambers from the perspective of model design will play a significant role for all kinds of users. 
The most straightforward way to mitigate echo chamber is to use a random recommendation policy. 
However, this is against the ultimate goal of recommendation, which is to accurately capture user preference and recommend items that user may like. 
As a result, it is impractical to give up the recommendation accuracy for the pursuit of mitigating echo chambers. Therefore, it is essential to mitigate echo chambers while maintaining the recommendation performance. In the next section, we will introduce our proposed framework for mitigating echo chambers.

\section{Methodology}\label{sec:model}
In this section, we will introduce our Dynamic Causal Collaborative Filtering ($\partial$CCF) framework in detail. The primary notations used throughout this section are detailed in Table \ref{tab:notation}.

\begin{table}[]
    \centering
    \begin{tabular}{c p{6.5cm}}
    \toprule
        \textbf{Symbol} & \textbf{Definition} \\\midrule
        $t$ & The timestamp \\
        $U$, $u$ & User variable and corresponding specific values\\
        $V_t$, $v_t$ & The variable denotes exposed item at time $t$ and the corresponding specific values\\
        $X_t$, $x_t$ & The variable denotes user history at time $t$ and the corresponding specific values\\
        $Y_t$, $y_t$ & The variable denotes the preference score at time $t$ and the corresponding specific values\\
        $x^*_t$ & The observed user history at time $t$ in real world\\
        $x'_t$ & The counterfactual user history at time $t$ which is unobserved in real world\\
        $v^*_t$ & The observed exposed item at time $t$ in real world\\
        $v'_t$ & The counterfactual exposed item at time $t$ which is unobserved in real world\\
        $y^*_t$ & The observed preference score at time $t$ in real world\\
        $y'_t$ & The counterfactual preference score at time $t$ which is unobserved in real world\\
        
         \bottomrule
    \end{tabular}
    \caption{Notations}
    \vspace{-20pt}
    \label{tab:notation}
\end{table}

\subsection{Preference Score Estimation}

Our causal graph in Figure \ref{fig:causalgraph}, with corresponding explanations in Section \ref{sec:causalgraph}, describes a dynamic data generation process in recommendation. Therefore, the proposed causal graph with loops can be temporally unfolded.
We capture a clip at time $t$ and time $t-1$ shown in Figure \ref{fig:causalgraph}(b). The unfolded causal graph represents the same data generation process as Figure \ref{fig:causalgraph}(a) and it helps to mitigate echo chambers in each step.

In this paper, we establish a counterfactual query similar to existing causal recommendation models \cite{wang2020causal,xiong2021counterfactual,wang2021counterfactual}, i.e., what would have happened if an item had been recommended, which can be mathematically represented as $P(y|u,do(v))$ \cite{xu2021causal}. In our case, we temporally unfold the causal graph and estimate the causal effect at each timestamp. 
Specifically, the desired estimation would be $P(y_t|u,do(v_t))$ at time $t$.
When we unfold the designed causal graph and capture a snapshot at time $t$ as in Figure \ref{fig:causalgraph}(b), we can apply the back-door adjustment to estimate the post-intervention effect based on the observational data. Since the variable set $\{U, X_t\}$ satisfies the back-door criterion for estimating $P(y_t|u,do(v_t))$ (i.e., variable $U$ blocks the back-door path $Y_t\leftarrow U\rightarrow V_t$ and variable $X_t$ blocks the back-door path $Y_t\leftarrow X_t\rightarrow V_t$), we are able to calculate $P(y_t|u,do(v_t))$ as follows.
\begin{equation}\label{eq:backdoor}
    P(y_t|u,do(v_t)) = \sum_{x_{t}} P(y_t|u,v_t,x_t)P(x_t|u)
\end{equation}

From Eq.\eqref{eq:backdoor}, the key difference between our causal model and the traditional associative models is the existence of conditional probability $P(x_t|u)$. 
In the real world, we denote the conditional probability $P(x^*_t|u)=1$ for an observed user history $x^*_t$ at time $t$ and $P(x'_t|u)=0$ for an unobserved user history $x'_t$ at time $t$.
However, observing user history $x^*_t$ does not imply that the user is destined to interact with items in $x^*_t$ at time $t$. Considering the counterfactual world, if a user had a chance to be recommended different items, they may also interact with those different items, thus $P(x'_t|u)$ is not necessarily to be zero. 
As a result, calculating Eq.\eqref{eq:backdoor} requires counterfactual reasoning beyond the observational data in the real world \cite{xu2021causal}. In the following, we will introduce how we can leverage counterfactual reasoning to estimate the preference score.

\subsection{Counterfactual Reasoning}
In our framework, the purpose of counterfactual reasoning is not only to enable the calculation of Eq.\eqref{eq:backdoor} but also to break the feedback loop in Figure \ref{fig:causalgraph} to mitigate echo chambers. 
As we mentioned before, the user may have a chance to interact with different items that belong to the unobserved history $x'_t$ at time $t$. 
Considering a record $(u,v_t,y_t)$ at time $t$ in the observational data, the user preference estimation $y_t=f_t(u,v)$ can be expressed in the following equations according to Eq.\eqref{eq:backdoor}.

\begin{equation}
\begin{aligned}
    y_t&=f_t(u,v) \propto P(y_t|u,do(v_t))\\
    &=\sum_{\tilde{x}_t} P(y_t|u,v_t,\tilde{x}_t)P(\tilde{x}_t|u)=E_{\tilde{x}_t|u}[P(y_t|u,v_t,\tilde{x}_t)]
\end{aligned}\label{eq:preference}
\end{equation}

Here we use $\tilde{x}_t$ to represent possible user histories, including the factual history $x^*_t$ and counterfactual histories $x'_t$. Based on Eq.\eqref{eq:preference}, the estimation for $P(y_t|u,do(v_t))$ is the expected estimation of $P(y_t|u,v_t,\tilde{x}_t)$.

\subsubsection{\textbf{Generate Counterfactual Histories}}\label{sec:ctfgeneration}
We will use the generated counterfactual histories to calculate Eq.\eqref{eq:preference} and mitigate echo chambers. In particular, we design a heuristic-based approach for counterfactual history generation.

First, we take a look at an example of how the historical sequence is generated in the feedback loop. 
Consider an observed status $(u,x^*_{t-1})$ for user $u$ at time $t-1$. Based on the observed history $x^*_{t-1}$, the system may recommend item $v^*_{t-1}$. If the user $u$ passively accepts the recommended item $v^*_{t-1}$ and likes it, then $v^*_{t-1}$ and $y^*_{t-1}=1$ would be a part of history $x^*_t$ at time $t$.
Based on our analysis in Section \ref{sec:echochambers}, such users will be affected by echo chambers if the recommendation model does not consider echo chamber. 
We cannot force users to actively explore different items in real-world recommender systems. Instead, we let users actively explore different items in the counterfactual world.
Following the principle of minimal changes in counterfactual reasoning \cite{wang2021counterfactual,tan2021counterfactual,xu2021causal,tan2022learning}, we replace the observed interaction at time $t-1$ to generate the counterfactual histories at time $t$:
\begin{equation}\label{eq:counterfactual}
    x'_t\leftarrow x^*_{t-1}, v'_{t-1}, y'_{t-1}
\end{equation}
Intuitively, this represents two possible situations: 1) the user still interacts with the recommended item $v^*_{t-1}$ but has a different preference $y'_{t-1}$ (i.e., $y'_{t-1}=0$ if $y^*_{t-1}=1$); 2) the user ignores item $v^*_{t-1}$ and interacts with a different item $v'_{t-1}$ and likes it (i.e., $y'_{t-1}=1$). To mitigate echo chambers, the user should interact with an item which is not similar with the recommended item $v^*_{t-1}$. We select counterfactual item $v'_{t-1}$ with least similarity to the recommended item $v^*_{t-1}$ to maximally mitigate echo chambers. 
\begin{equation}\label{eq:ctfitem}
    v'_{t-1} = \argmin_v Similarity(v, v^*_{t-1})
\end{equation}

Here the similarity is calculated by the dot product between the embedding of the two items. It is worth mentioning that the counterfactual item $v'_{t-1}$ is not a real interaction nor an item to be recommended in reality. 

\subsubsection{\textbf{Calculate the Expectation}}
Assume that we have generated $n$ counterfactual histories $\{x'^{(1)}_t, \cdots, x'^{(n)}_t\}$ for user $u$ at time $t$. Then, we can calculate $P(y_t|u,do(v_t))$ based on the factual history $x^*_t$ and the counterfactual histories $\{x'^{(i)}_t\}_{i=1}^{n}$ according to Eq.\eqref{eq:preference}. For simplicity, we consider $P(\tilde{x}_t|u)$ as a piece-wise uniform distribution over the factual and counterfactual histories, i.e., 
\begin{equation}\label{eq:uniform}
P(\tilde{x}_t|u) = 
\begin{cases} 
\alpha,~ \text{when}~\tilde{x}_t = x^*_t \\ 
\beta,~ \text{when}~\tilde{x} = x'^{(i)}_t,~ i\in\{1,2\cdots n\}
\end{cases},
\alpha+n\beta=1
\end{equation}
where $\alpha$ is the probability of the factual example $x^*_t$, and $\beta$ is the probability of each counterfactual example $x'^{(i)}_t$. Since $x^*_t$ is already observed, we apply a higher probability to $x^*_t$ than $x'^{(i)}_t$, i.e., $\alpha>\beta>0$.
Then we have:
\begin{equation}
\begin{aligned}
\label{eq:expectation}
    P(y_t|u,do(v_t)) &= \sum_{\tilde{x}_t} P(y_t|u,v_t,\tilde{x}_t)P(\tilde{x}_t|u)\\ 
    &= \alpha~ P_g(y_t|u,v_t, x^*_t)+\beta \sum_{i=1}^n P_g(y_t|u,v_t,x'^{(i)}_t)
\end{aligned}
\end{equation}
We use $P_g$ to represent the probability estimation of the base recommendation algorithm.

As we mentioned in Section \ref{sec:ctfgeneration}, the counterfactual histories intuitively represent the possible historical records if a user ignores the recommended items. Additionally, if we treat the factual history as a result of following recommender systems, the probability $\alpha$ in Eq.\eqref{eq:expectation} can be considered as sharing the same meaning with probability $p$ in Eq.\eqref{eq:transition3}. As we mentioned in Section \ref{sec:echochambers}, if the probability $p$ is properly chosen, users will benefit from personalized recommendation while avoiding homogeneity brought by echo chambers. We will explore the effect of probability $\alpha$ in Section \ref{sec:sensitivity}. 

\section{Experiments}\label{sec:experiments}
In this section, we will first describe the evaluation metrics, datasets, baselines and implementation details and then provide our results and discussion.

\begin{table}[t]
    \centering
    \caption{The Statistic of the Datasets}
    \vspace{-10pt}
    \begin{tabular}{c|crrr}
    \toprule
        Dataset & Phase & \# Users & \# Items & \# Interactions \\\midrule
        \multirowcell{4}{Movielens-1m} & All & 6,040 & 3,706 & 1,000,209 \\
        & Phase1 & 1,611 & 3,322 & 226,543 \\
        & Phase2 & 2,878 & 3,505 & 382,025 \\
        & Phase3 & 2,738 & 3,627 & 391,641 \\\midrule
        \multirowcell{4}{Electronics} & All & 33,602 & 16,448 & 788,143 \\
        & Phase1 & 8,495 & 3,679 & 39,489 \\
        & Phase2 & 29,798 & 13,820 & 377,460 \\
        & Phase3 & 31,193 & 15,162 & 371,194 \\
        \bottomrule
    \end{tabular}
    \label{tab:datasets}
    \vspace{-10pt}
\end{table}

\begin{table*}[t]
    \centering
    \footnotesize
    \setlength{\tabcolsep}{2.pt}
    \resizebox{1.0\textwidth}{!}{
    \begin{tabular}{c|c|cccccccccc|cccccccccc}
    \toprule
    \multicolumn{2}{c}{\multirow{3}{*}{Methods}} & \multicolumn{10}{c}{Movielens 1m} & \multicolumn{10}{c}{Electronics} \\\cmidrule(lr){3-12}\cmidrule(lr){13-22}
    \multicolumn{2}{c}{} & \multicolumn{4}{c}{Phase2} & \multicolumn{4}{c}{Phase3} & \multicolumn{2}{c}{Echo Chambers} & \multicolumn{4}{c}{Phase2} & \multicolumn{4}{c}{Phase3} & \multicolumn{2}{c}{Echo Chambers}\\\cmidrule(lr){3-6}\cmidrule(lr){7-10}\cmidrule(lr){11-12}\cmidrule(lr){13-16}\cmidrule(lr){17-20}\cmidrule(lr){21-22}
    \multicolumn{2}{c}{} & n@10 & H@10 & un@10 & uH@10 & n@10 & H@10 & un@10 & uH@10 & $\Delta 12$ & $\Delta 13$ & n@10 & H@10 & un@10 & uH@10 & n@10 & H@10 & un@10 & uH@10 & $\Delta 12$ & $\Delta 13$\\\midrule
    \multirow{7}{*}{GRU4Rec} & Original & 0.5293 & 0.7797 & 0.1939 & 0.3073 & 0.5072 & 0.7482 & 0.1848 & 0.2929 & 0.996 & 1.174 & 0.2726 & 0.4567 & 0.0816 & 0.1294 & 0.3206 & 0.5400 & 0.0942 & 0.1494 & 6.270 & 7.205\\
     & MMR & 0.5267 & 0.7747 & 0.1925 & 0.3052 & \textbf{0.5102} & 0.7467 & \underline{\textbf{0.1900}} & \underline{\textbf{0.3012}} & \textbf{0.971} & 1.179 & 0.2614 & 0.4374 & \textbf{0.0820} & 0.1274 & 0.3156 & 0.5304 & 0.0939 & \textbf{0.1495} & \textbf{6.138} & \textbf{7.178}\\
     & IPS & 0.5233 & 0.7640 & 0.1919 & 0.3041 & 0.5001 & 0.7452 & 0.1786 & 0.2831 & \textbf{0.956} & \textbf{1.079} & 0.2678 & \textbf{0.4582} & 0.0788 & 0.1290 & 0.3183 & 0.5376 & 0.0909 & 0.1483 & 6.783 & 7.275\\
     & CausE & \textbf{0.5310} & \textbf{0.7836} & \textbf{0.1954} & \textbf{0.3107} & 0.5058 & \underline{\textbf{0.7546}} & 0.1837 & \textbf{0.2952} & 1.244 & 1.352 & \textbf{0.2764} & \textbf{0.4617} & \textbf{0.0829} & \textbf{0.1315} & 0.3123 & 0.5339 & 0.0891 & 0.1412 & 6.395 & \underline{\textbf{7.154}}\\
     & CCF & \textbf{0.5327} & \textbf{0.7807} & \textbf{0.1972} & \textbf{0.3098} & \underline{\textbf{0.5091}} & \textbf{0.7527} & \textbf{0.1865} & \textbf{0.2937} & 1.262 & 1.342 & \textbf{0.2838} & \textbf{0.4708} & \textbf{0.0846} & \textbf{0.1367} & \underline{\textbf{0.3276}} & \underline{\textbf{0.5427}} & \underline{\textbf{0.0952}} & 0.1490 & 6.832 & 7.421\\
     & DICE & \underline{\textbf{0.5372}} & \underline{\textbf{0.7882}} & \textbf{0.1986} & \textbf{0.3148} & 0.5026 & 0.7410 & 0.1822 & 0.2887 & 1.244 & 1.349 & \underline{\textbf{0.2862}} & \underline{\textbf{0.4727}} & \underline{\textbf{0.0885}} & \underline{\textbf{0.1402}} & 0.3178 & 0.5320 & \textbf{0.0943} & \textbf{0.1495} & 7.052 & 7.630\\
     & MACR & \textbf{0.5360} & \textbf{0.7815} & \underline{\textbf{0.2009}} & \underline{\textbf{0.3184}} & 0.5067 & 0.7441 & 0.1834 & 0.2906 & 1.167 & 1.255 & 0.2699 & 0.4511 & 0.0808 & 0.1280 & 0.3200 & 0.5377 & \textbf{0.0947} & \underline{\textbf{0.1500}} & \textbf{6.163} & 7.218\\
     & $\partial$CCF & 0.5268 & 0.7772 & 0.1817 & \textbf{0.3081} & 0.4999 & \textbf{0.7494} & 0.1848 & 0.2912 & \underline{\textbf{0.743}} & \underline{\textbf{0.979}} & \textbf{0.2761} & \textbf{0.4615} & \textbf{0.0826} & \textbf{0.1308} & \textbf{0.3227} & 0.5351 & 0.0903 & 0.1475 & \underline{\textbf{5.388}} & \textbf{7.199}\\\midrule
     
    \multirow{7}{*}{STAMP} & Original & 0.5002 & 0.7401 & 0.1813 & 0.2873 & 0.4776 & 0.7214 & 0.1681 & 0.2665 & 2.120 & 2.244 & 0.2839 & 0.4687 & 0.0878 & 0.1391 & 0.3065 & 0.5206 & 0.0882 & 0.1398 & 3.910 & 4.152\\
     & MMR & 0.4845 & 0.7216 & 0.1784 & 0.2827 & 0.4667 & 0.6962 & \textbf{0.1705} & \textbf{0.2703} & \textbf{2.118} & \textbf{2.240} & 0.2788 & 0.4594 & \textbf{0.0883} & \textbf{0.1399} & 0.3025 & 0.5138 & 0.0873 & 0.1383 & 3.923 & \textbf{4.133}\\
     & IPS & 0.4920 & 0.7401 & 0.1743 & 0.2763 & 0.4761 & 0.7199 & 0.1679 & 0.2661 & \underline{\textbf{1.779}} & \underline{\textbf{1.882}} & \textbf{0.2908} & \textbf{0.4764} & \textbf{0.0903} & \textbf{0.1431} & \underline{\textbf{0.3258}} & \underline{\textbf{0.5483}} & \textbf{0.0952} & \textbf{0.1508} & 3.933 & 4.318\\
     & CausE & 0.4999 & \textbf{0.7483} & \textbf{0.1847} & \textbf{0.2885} & \textbf{0.4816} & \textbf{0.7267} & 0.1667 & \textbf{0.2705} & \textbf{2.116} & \textbf{2.231} & 0.2828 & \textbf{0.4754} & 0.0878 & \underline{\textbf{0.1458}} & \textbf{0.3172} & \textbf{0.5402} & \textbf{0.0928} & \textbf{0.1411} & 3.934 & 4.237\\
     & CCF & \textbf{0.5037} & \textbf{0.7412} & 0.1802 & 0.2867 & \textbf{0.4815} & \textbf{0.7260} & \textbf{0.1683} & \textbf{0.2691} & \textbf{2.112} & 2.249 & \underline{\textbf{0.2926}} & \underline{\textbf{0.4795}} & \textbf{0.0901} & \textbf{0.1428} & \textbf{0.3247} & \textbf{0.5440} & \underline{\textbf{0.0963}} & \underline{\textbf{0.1527}} & 3.933 & 4.254\\
     & DICE & \underline{\textbf{0.5080}} & \underline{\textbf{0.7612}} & \underline{\textbf{0.1879}} & \underline{\textbf{0.3011}} & \underline{\textbf{0.4868}} & \underline{\textbf{0.7295}} & \underline{\textbf{0.1742}} & \underline{\textbf{0.2801}} & 2.187 & 2.295 & 0.2818 & 0.4618 & 0.0865 & 0.1343 & 0.2909 & 0.5024 & 0.0802 & 0.1272 & 3.918 & 4.205\\
     & MACR & \textbf{0.5023} & \textbf{0.7462} & \textbf{0.1858} & \textbf{0.2945} & 0.4746 & 0.7158 & \textbf{0.1710} & \textbf{0.2710} & \textbf{2.082} & \textbf{2.203} & \textbf{0.2912} & \textbf{0.4766} & \underline{\textbf{0.0910}} & \textbf{0.1442} & \textbf{0.3082} & \textbf{0.5239} & 0.0881 & 0.1396 & \textbf{3.903} & 4.177\\
     & $\partial$CCF & 0.4962 & \textbf{0.7444} & 0.1803 & 0.2842 & 0.4720 & 0.7192 & 0.1658 & 0.2627 & \textbf{2.111} & \textbf{2.242} & \textbf{0.2849} & \textbf{0.4756} & \textbf{0.0891} & 0.1349 & \textbf{0.3189} & \textbf{0.5409} & \textbf{0.0929} & \textbf{0.1472} & \underline{\textbf{3.901}} & \underline{\textbf{4.132}}\\\midrule
     
    \multirow{7}{*}{NCR} & Original & 0.4853 & 0.7579 & 0.1597 & 0.2531 & 0.4725 & 0.7331 & 0.1572 & 0.2492 & 1.028 & 1.074 & 0.2522 & 0.4287 & 0.0722 & 0.1145 & 0.2737 & 0.4739 & 0.0750 & 0.1189 & 4.705 & 5.002\\
     & MMR & 0.4853 & 0.7579 & \textbf{0.1633} & \textbf{0.2588} & 0.4560 & 0.7218 & 0.1470 & 0.2329 & \textbf{0.953} & \textbf{1.027} & \textbf{0.2549} & \underline{\textbf{0.4363}} & \textbf{0.0723} & 0.1145 & 0.2720 & 0.4734 & 0.0749 & 0.1187 & \textbf{4.676} & \textbf{4.918}\\
     & IPS & 0.4833 & \textbf{0.7622} & 0.1561 & 0.2474 & 0.4626 & \underline{\textbf{0.7410}} & 0.1427 & 0.2262 & \textbf{1.003} & 1.077 & \textbf{0.2596} & \textbf{0.4345} & \textbf{0.0773} & \textbf{0.1237} & \textbf{0.2845} & \textbf{0.4887} & \underline{\textbf{0.0796}} & \underline{\textbf{0.1261}} & \textbf{4.600} & \textbf{4.920} \\
     & CausE & \underline{\textbf{0.4959}} & \underline{\textbf{0.7676}} & \textbf{0.1637} & \textbf{0.2613} & \underline{\textbf{0.4794}} & \textbf{0.7369} & \underline{\textbf{0.1685}} & \textbf{0.2533} & 1.059 & 1.105 & \textbf{0.2562} & \textbf{0.4327} & \textbf{0.0766} & \textbf{0.1217} & \textbf{0.2780} & \textbf{0.4859} & \textbf{0.0774} & \textbf{0.1241} & \textbf{4.595} & \textbf{4.937}\\
     & CCF & 0.4824 & 0.7533 & \textbf{0.1603} & 0.2497 & \textbf{0.4751} & \textbf{0.7403} & \textbf{0.1598} & 0.2478 & \textbf{1.015} & 1.087 & 0.2521 & \textbf{0.4306} & \textbf{0.0754} & \textbf{0.1227} & \underline{\textbf{0.2853}} & \underline{\textbf{0.4929}} & \textbf{0.0792} & \textbf{0.1255} & \textbf{4.486} & \textbf{4.993}\\
     & DICE & \textbf{0.4868} & 0.7526 & 0.1579 & 0.2501 & 0.4714 & 0.7275 & 0.1537 & 0.2473 & 1.032 & \textbf{1.050} & \textbf{0.2590} & \textbf{0.4346} & \textbf{0.0769} & \textbf{0.1235} & \textbf{0.2754} & \textbf{0.4791} & \textbf{0.0759} & 0.1174 & \textbf{4.650} & \textbf{4.925}\\
     & MACR & \textbf{0.4936} & \textbf{0.7615} & \underline{\textbf{0.1640}} & \textbf{0.2599} & \textbf{0.4746} & 0.7233 & \textbf{0.1631} & \underline{\textbf{0.2586}} & 1.043 & \textbf{1.060} & \textbf{0.2592} & \textbf{0.4342} & \textbf{0.0775} & \textbf{0.1228} & \textbf{0.2757} & \textbf{0.4798} & \textbf{0.0753} & \textbf{0.1193} & \textbf{4.488} & \textbf{4.965}\\
     & $\partial$CCF  & \textbf{0.4941} & 0.7547 & 0.1590 & \underline{\textbf{0.2724}} & 0.4659 & 0.7162 & \textbf{0.1603} & \textbf{0.2541} & \underline{\textbf{0.936}} & \underline{\textbf{0.985}} & \underline{\textbf{0.2614}} & \textbf{0.4362} & \underline{\textbf{0.0784}} & \underline{\textbf{0.1242}} & \textbf{0.2781} & \textbf{0.4762} & \textbf{0.0787} & \textbf{0.1248} & \underline{\textbf{4.323}} & \underline{\textbf{4.841}}\\\bottomrule
    \end{tabular}
    }
    \caption{Overall performance of applying our framework on three recommendation models. The recommendation performance is evaluated as a ranking task. n, H, un, uH represent nDCG, Hit, u$\_$nDCG and u$\_$Hit respectively. The effect of echo chambers is measured by change of content diversity (the lower the better). We calculate the content diversity at each phase and report the change of content diversity. We use $\Delta ij$ to denote the change of content diversity between phase $i$ and phase $j$. The improved performance is bold and the best is underlined. 
    }
    \vspace{-15pt}
    \label{tab:results}
\end{table*}

\subsection{Evaluation of Echo Chamber Effects}\label{sec:echochambereval}
Echo chambers will result in homogeneity and polarization, hence over time the system will narrow the user exposure towards some specific content. 
Therefore, similar to \cite{ge2020understanding}, we evaluate the effect of echo chambers by measuring the changes of content diversity.

According to \cite{kaminskas2016diversity,ge2020understanding}, we use the average of the pairwise distance (i.e., Euclidean distance between item embeddings) to measure the content diversity, and use the temporal changes of content diversity to measure the echo chamber effect of the recommender systems.

\subsection{Dataset Description}
Our experiments are conducted on two real-world datasets from \textit{Amazon}\footnote{https://nijianmo.github.io/amazon/} \cite{ni2019justifying} and \textit{Movielens}\footnote{https://grouplens.org/datasets/movielens/} \cite{harper2015movielens}. More specifically, we use \textit{Electronics} category from Amazon and \textit{Movielen-1m} from Movielens. 

In order to measure the effect of echo chambers by the temporal change of content diversity, we use two timestamps to chronologically split the data into three phases. 
The \textit{Movielens-1m} dataset contains the data in 2000, so we split the data at July 31th, 2000 and November 20th, 2000. The \textit{Electronics} dataset contains data spanning May 1996 to October 2018. We use December 31th, 2010 and June 30th, 2015 as two timestamps. The statistics of the datasets are summarized in Table \ref{tab:datasets}.

For all datasets, following prior works,
we consider ratings $\geq 4$ as positive feedback (likes) and ratings $\leq 3$ as negative feedback (dislikes). Meanwhile, for all three phases in two datasets, we apply leave-one-out to split training, validation and test data \cite{zhao2020revisiting}.

\subsection{Baseline Models}
We also employ other six frameworks for comparison, including one re-ranking framework and five causal learning frameworks. 
\textbf{MMR}~\cite{carbonell1998use} is a re-ranking model for information retrieval which tries to maximize the relevance and novelty in finally retrieved top-ranked items.
\textbf{IPS}~\cite{schnabel2016recommendations} is an Inverse Propensity Scoring-based model, which uses a propensity estimator to re-weight the training samples to eliminate popularity bias. More specifically, we apply the clipped propensity score as \cite{saito2020unbiased} to reduce the variance of IPS.
\textbf{CausE} ~\cite{bonner2018causal} creates two sets of embeddings for unbiased and biased data separately and regularization is applied to force the two sets of embeddings similar.
\textbf{CCF} ~\cite{xu2021causal} is a causal collaborative filtering framework which uses counterfactual histories and applies counterfactual constraints to estimate causal preference.
\textbf{DICE} ~\cite{zheng2021disentangling} is a framework for popularity bias problem, which disentangles user interest and conformity into two sets of embeddings.
\textbf{MACR} ~\cite{wei2021model} is a model-agnostic framework for alleviating popularity bias issue in recommender systems, which performs multi-task training according to causal graph to assess the contribution of different causes on the ranking score. Finally, we use \textbf{$\bm\partial$CCF} to denote our dynamic causal collaborative filtering framework.

We apply all above frameworks on three recommendation models, including two sequential models and a reasoning model. 
\textbf{GRU4Rec} \cite{hidasi2015session} is a session-based recommendation model, which uses recurrent neural networks---in particular, Gated Recurrent Units (GRU)---to capture sequential patterns.
\textbf{STAMP} \cite{liu2018stamp} is the Short-Term Attention/Memory Priority model, which takes attention mechanism to model both short-term and long-term user preferences.
\textbf{NCR} \cite{chen2021neural} is the Neural Collaborative Reasoning model, which employs neural logic reasoning for recommendation in a logical space.

\subsection{Implementation Details}
For recommendation performance, we evaluate models as a top-$K$ recommendation task. For each user in the validation and test set, we randomly sample 100 negative items for ranking evaluation, where the negative items are either negative feedback items (i.e., items that user dislikes) or non-interacted items. The models are evaluated on four metrics, two are nDCG@10 and Hit@10 metrics calculated by ranking on sampled testing data, and the other two are nDCG@10 and Hit@10 metrics calculated by corrected rank as \cite{krichene2020sampled} to get an unbiased evaluation (denote as u$\_$nDCG@10 and u$\_$Hit@10). For performance of mitigating echo chambers, we evaluate the change of content diversity (i.e., introduced in Section \ref{sec:echochambereval}) based on the list of recommendations with length 10.

For the hyper-parameters, we take the embedding dimension as 64, the structure of neural networks is a two-layer MLP with dimension 64. We set the learning rate to 0.001 and the  $\ell_2$-regularization weight to 1e-6. For all base models, we set the maximum length of history as 10. For our framework, we generate 9 counterfactual histories for each positive user-item pair (i.e., $n$ in Eq.\eqref{eq:expectation}) and set the factual probability $\alpha$ as 0.3. We will discuss the influence of these two parameters in Section \ref{sec:sensitivity}. We report the performance of each model with the best performance on validation data based on nDCG@10. For all base models, We first train the base model on phase 1 data to make sure the content diversity at phase 1 is consistent before and after applying the frameworks. Then, on phase 2 and phase 3 data, we separately train the base model with and without frameworks to measure the changes of content diversity. 


\begin{figure*}[t!]
\captionsetup[sub]{font=small,labelfont=normalfont,textfont=normalfont}
    \centering
    \begin{subfigure}{0.24\textwidth}
        \includegraphics[scale=0.31]{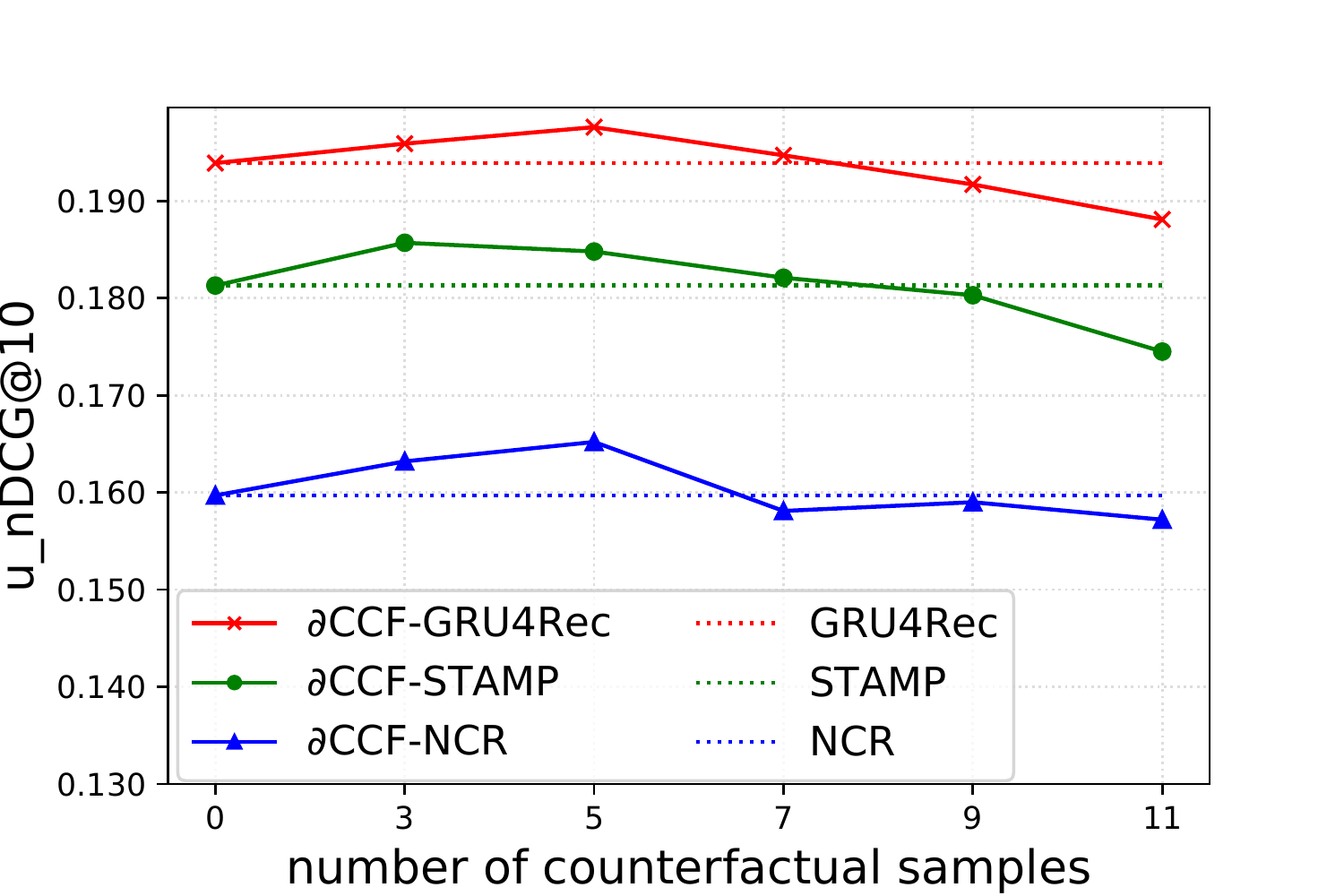}
        \caption{Movielen-1m (u$\_$nDCG@10)}
    \end{subfigure}
    \begin{subfigure}{0.24\textwidth}
        \includegraphics[scale=0.31]{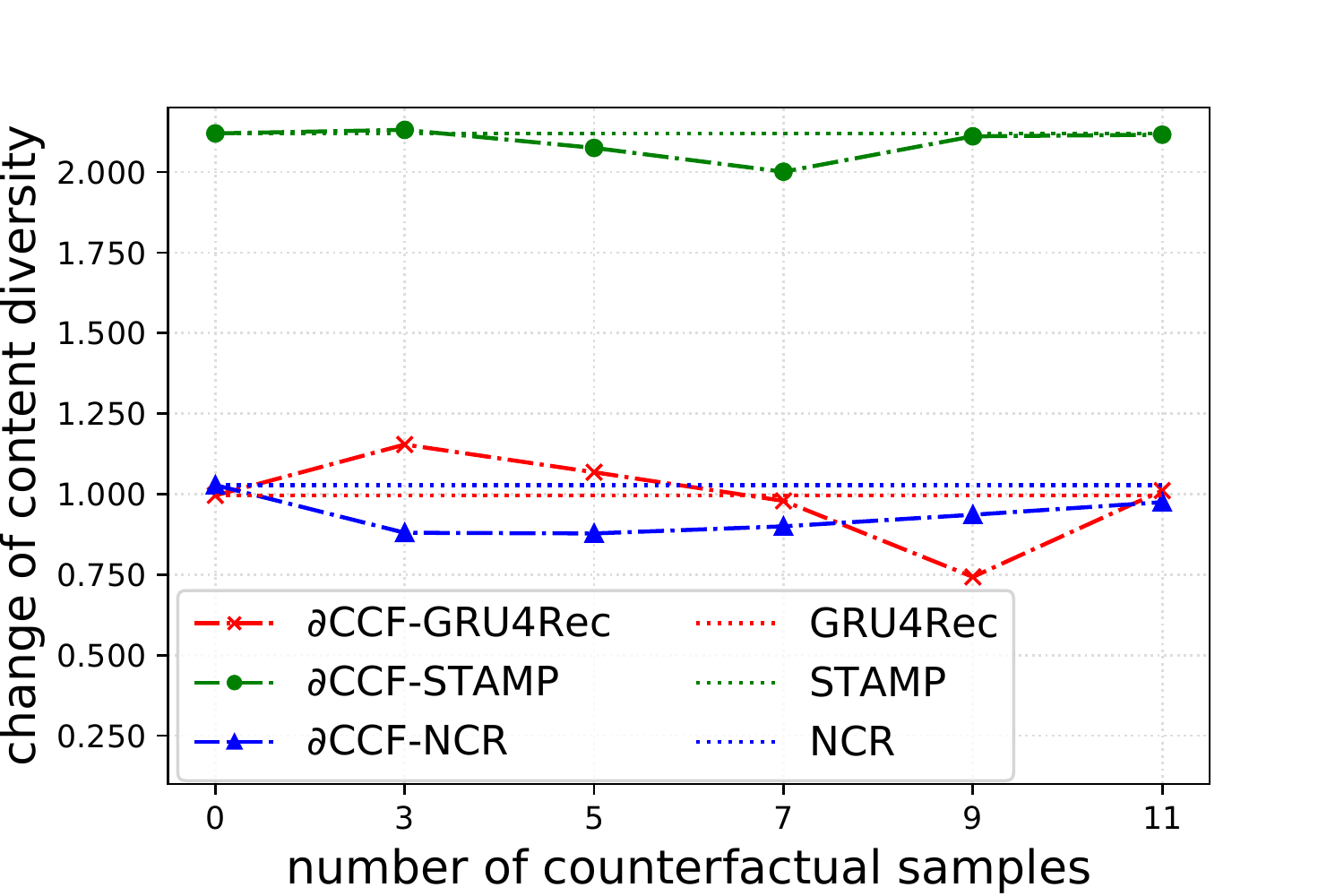}
        \caption{Movielen-1m ($\Delta 12$)}
    \end{subfigure}
    \begin{subfigure}{0.24\textwidth}
        \includegraphics[scale=0.31]{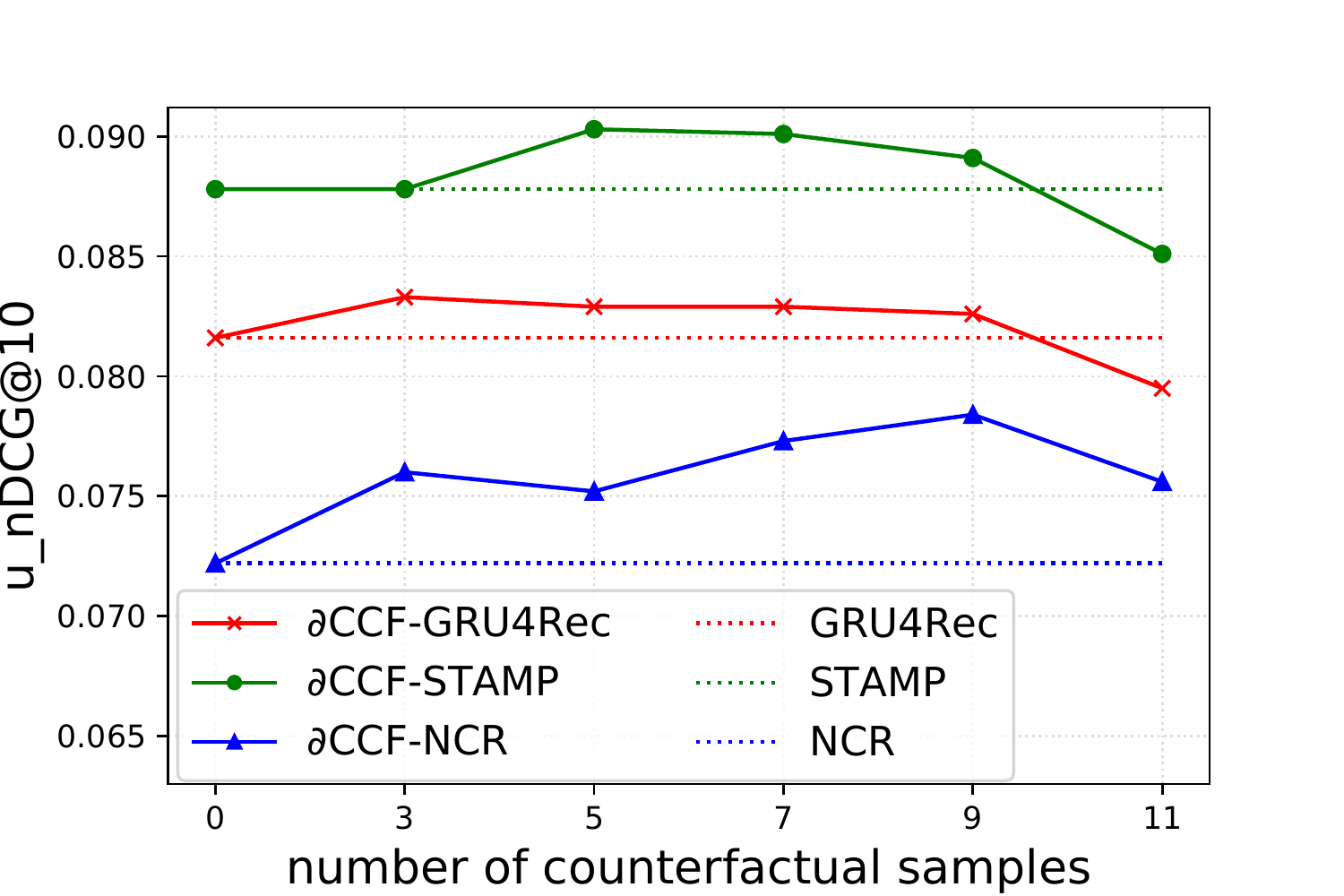}
        \caption{Electronics (u$\_$nDCG@10)}
    \end{subfigure}
    \begin{subfigure}{0.24\textwidth}
        \includegraphics[scale=0.31]{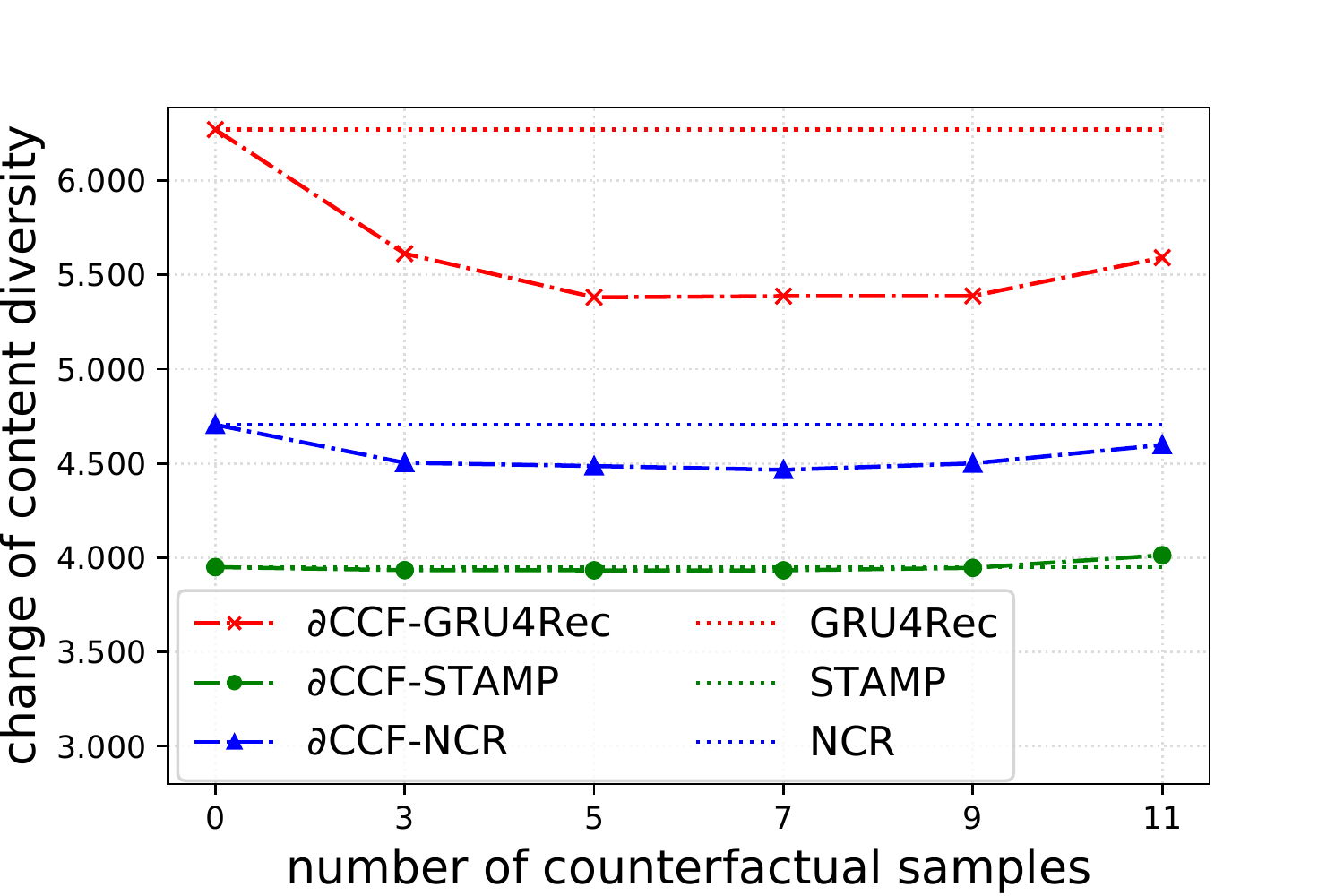}
        \caption{Electronics ($\Delta 12$)}
    \end{subfigure}
    \vspace{-5pt}
    \caption{The influence of different number of counterfactual samples ($n$ in Eq.\eqref{eq:expectation}). (a) and (c) are recommendation performance (u$\_$nDCG@10) on phase 2 of the two datasets. (b) and (d) are the change of content diversity ($\Delta 12$ in Table \ref{tab:results}) on phase 2 of the two datasets. The change of content diversity is the lower the better.}
    \label{fig:ctf_num}
    \vspace{-10pt}
\end{figure*}

\begin{figure*}[t!]
\captionsetup[sub]{font=small,labelfont=normalfont,textfont=normalfont}
    \centering
    \begin{subfigure}{0.24\textwidth}
        \includegraphics[scale=0.31]{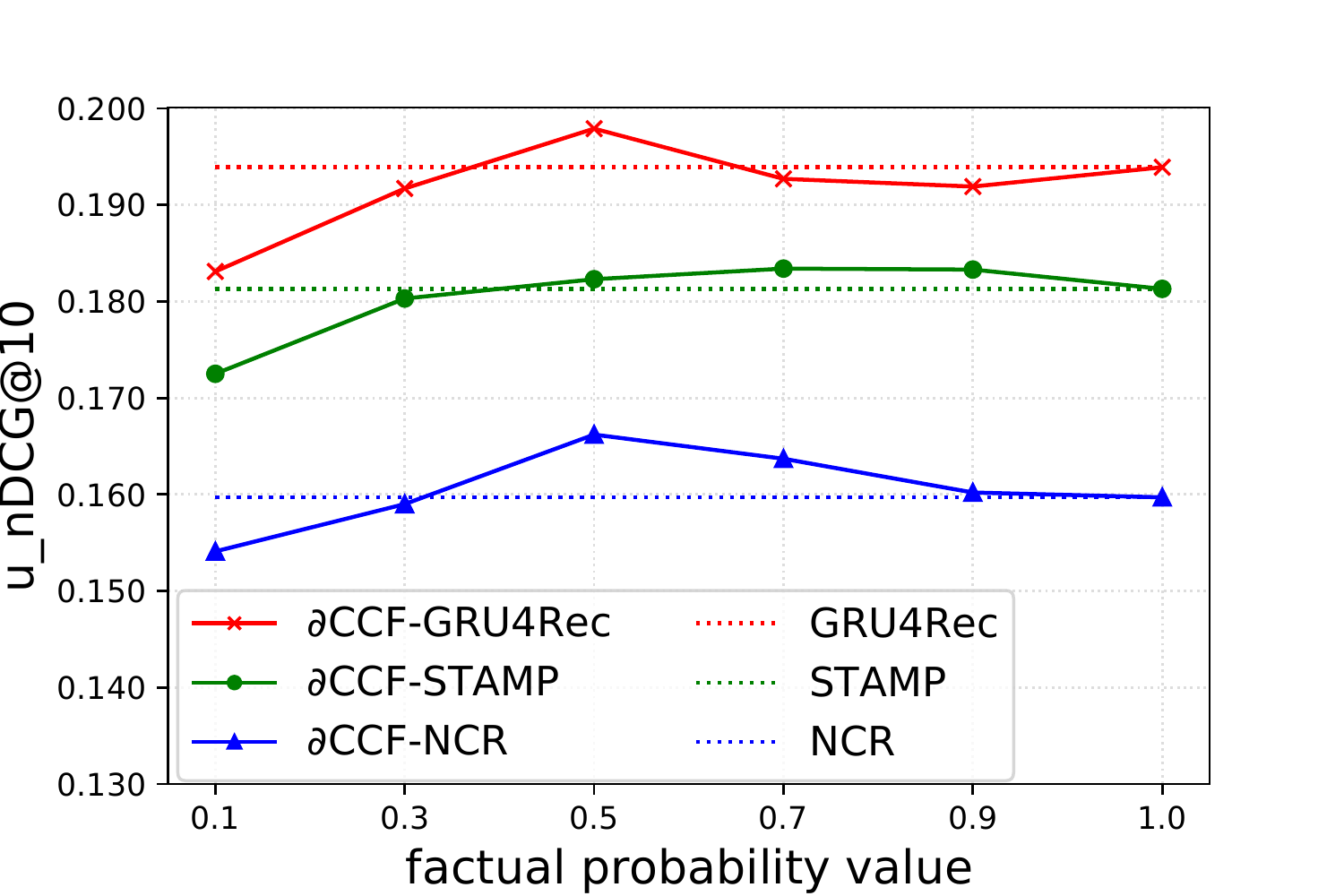}
        \caption{Movielen-1m (u$\_$nDCG@10)}
    \end{subfigure}
    \begin{subfigure}{0.24\textwidth}
        \includegraphics[scale=0.31]{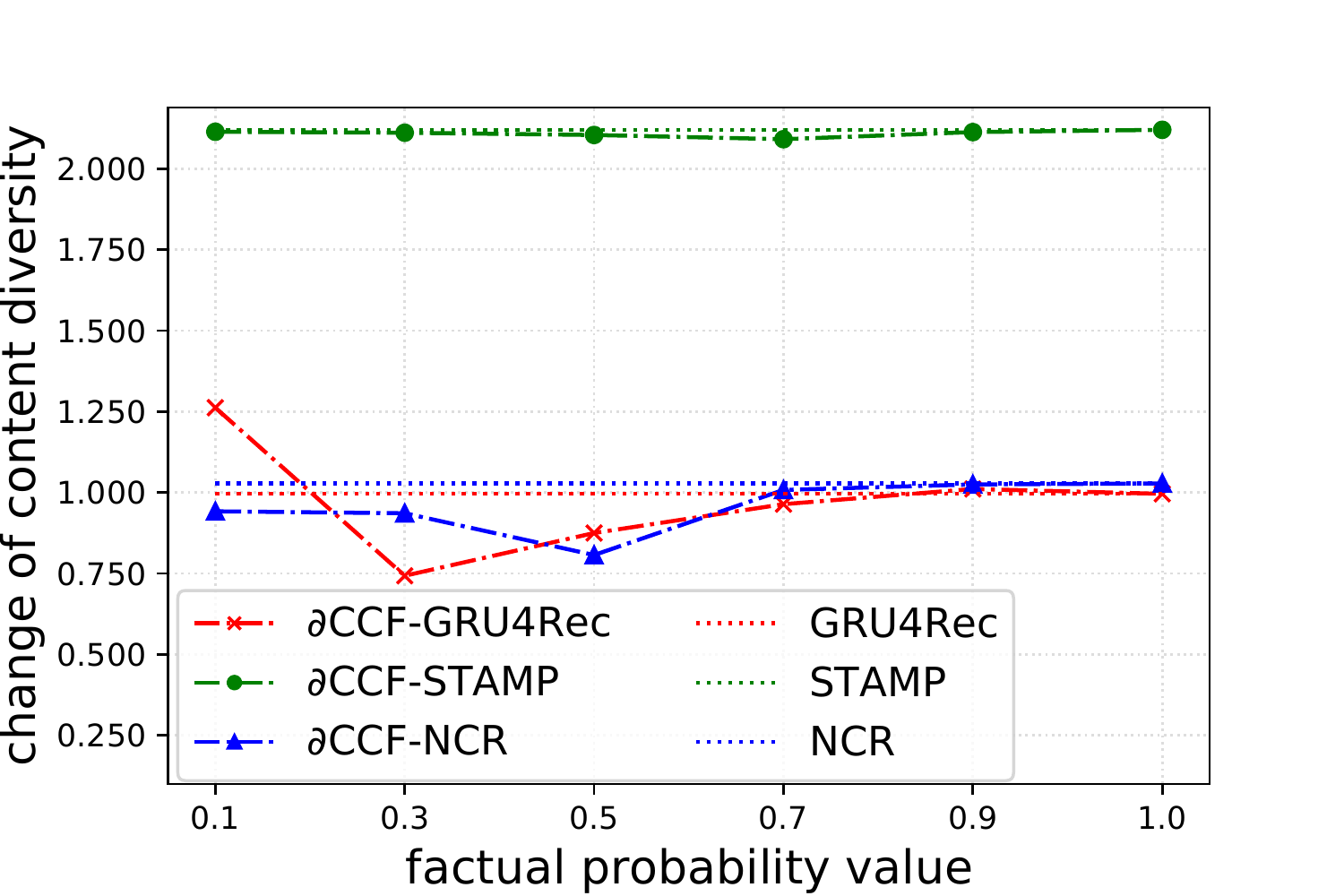}
        \caption{Movielen-1m ($\Delta 12$)}
    \end{subfigure}
    \begin{subfigure}{0.24\textwidth}
        \includegraphics[scale=0.31]{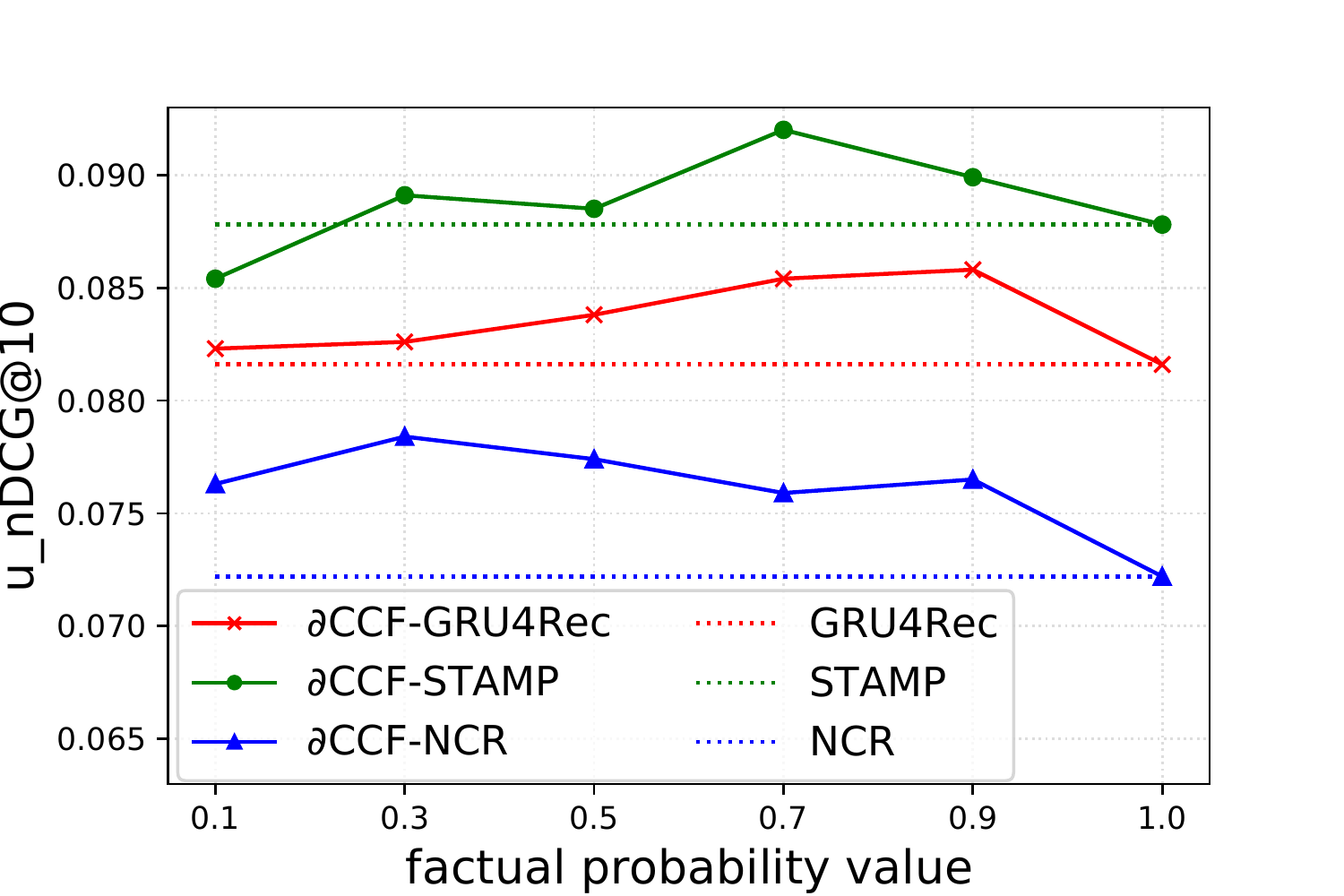}
        \caption{Electronics (u$\_$nDCG@10)}
    \end{subfigure}
    \begin{subfigure}{0.24\textwidth}
        \includegraphics[scale=0.31]{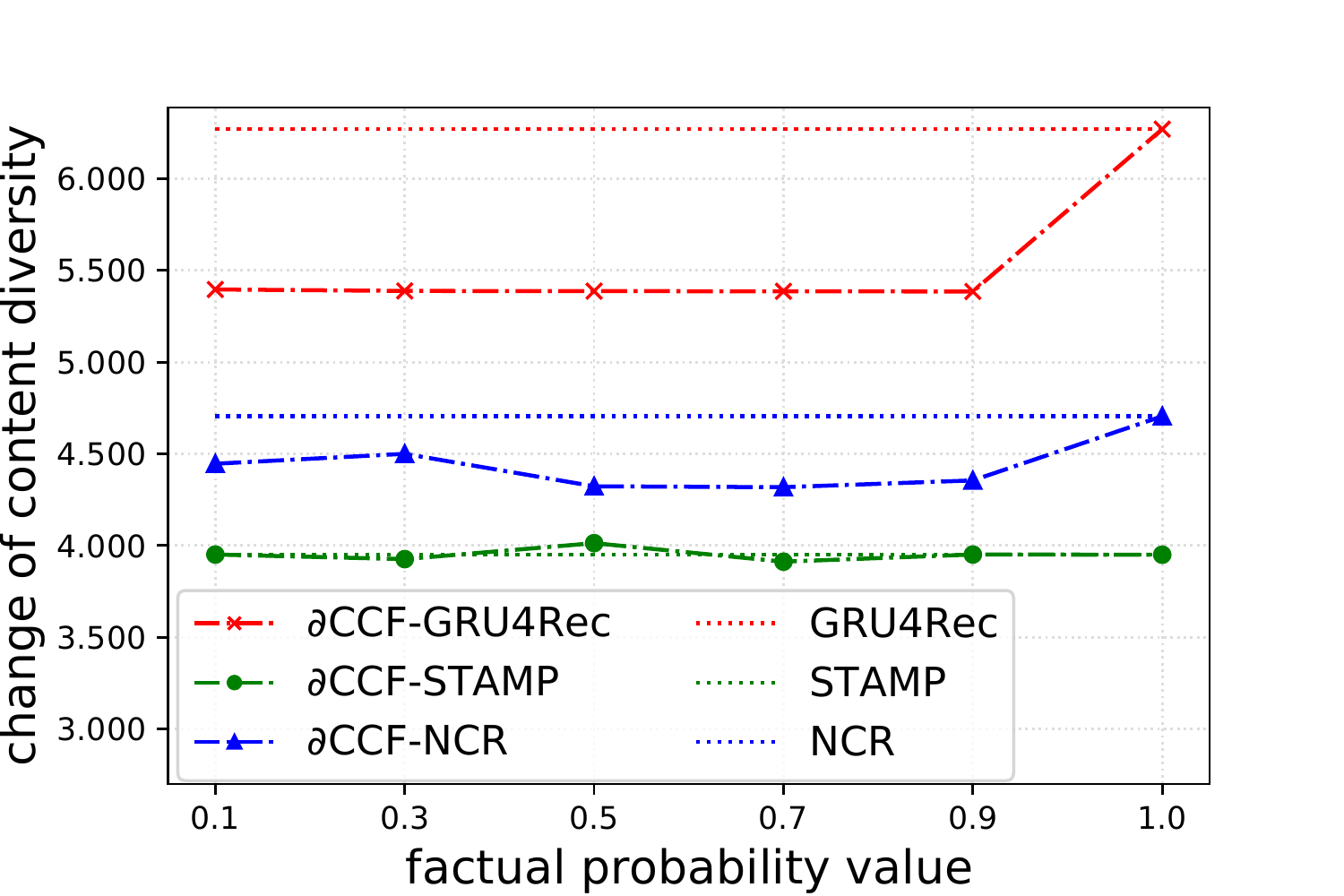}
        \caption{Electronics ($\Delta 12$)}
    \end{subfigure}
    \vspace{-5pt}
    \caption{The influence of different factual probability ($\alpha$ in Eq.\ref{eq:expectation}). (a) and (c) are recommendation performance (u$\_$nDCG@10) on phase 2 of the two datasets. (b) and (d) are the change of content diversity ($\Delta 12$ in Table \ref{tab:results}) on phase 2 of the two datasets. The change of content diversity is the lower the better.}
    \label{fig:fact_prob}
    \vspace{-10pt}
\end{figure*}

\subsection{Results and Discussions}
In Table \ref{tab:results}, we present the overall performance of applying our framework and baseline frameworks on three base recommendation models over two datasets, including recommendation performance (nDCG@10, Hit@10, u$\_$nDCG@10 and u$\_$Hit@10) and echo chambers evaluation (the changes of content diversity).

From Table \ref{tab:results}, we can observe that the effects of echo chambers have been mitigated (i.e., the change of content diversity has been reduced) after applying our $\partial$CCF framework for all three base models over two datasets. Averaging the change of content diversity across all three base models over two phases of both datasets, the change of content diversity has been reduced by 7.2\%. The largest improvement 25.4\% is obtained at phase 2 of the Movielens-1m dataset after applying on GRU4Rec. 
For the recommendation performance, we can observe that applying our framework over the base models will not hurt the recommendation performance. We can achieve comparable recommendation performance after applying our framework. When averaging across all recommendation metrics on all datasets and base models, our $\partial$CCF framework achieves 1\% improvement on recommendation performance. The largest improvement is 8.6\% on u$\_$nDCG@10 
of NCR at phase2 of the Electronics dataset. Therefore, our framework is capable of mitigating the echo chambers without sacrificing much recommendation performance.

The baseline frameworks include five causal learning frameworks (i.e., IPS, CausE, CCF, DICE, MACR) and one re-ranking framework (i.e., MMR). For the five causal learning frameworks, the goals of these causal learning frameworks are improving recommendation performance. Therefore, the best recommendation performance (i.e., the underlined values on recommendation metrics in Table \ref{tab:results}) are usually obtained by those causal learning frameworks. Although these causal learning baseline frameworks can be helpful in mitigating echo chambers in some cases, the best performance on mitigating echo chambers is provided by our $\partial$CCF framework in most cases. Among the 2 (datasets) $\times$ 2 (phases) $\times$ 3 (base models) = 12 cases, the $\partial$CCF framework achieves 9 best performance on mitigating echo chambers. For the re-ranking framework MMR, it tries to maximize both the relevance and the novelty, which is an accuracy-diversity trade-off. According to Table \ref{tab:results}, our $\partial$CCF framework achieves better performance of mitigating echo chambers (i.e., achieves 9/12 best performance on mitigating echo chambers). Comparing the recommendation performance with the base model, MMR decreases the recommendation performance by 1\% when averaging across all recommendation metrics on all datasets and base models. Compared with MMR, $\partial$CCF can incorporate counterfactual reasoning to avoid hurting the recommendation performance.

In summary, our framework gets better performance than other baseline frameworks on mitigating echo chambers without sacrificing the recommendation performance in most cases.

\subsection{User Satisfaction}
One consequence of echo chambers is gradually hurting user's satisfaction. Following \cite{gao2022cirs}, we evaluate models with the cumulative satisfaction to show the performance of mitigating echo chambers. Specifically, user satisfaction is defined as:
\begin{equation}
    \text{Satisfaction}=
    \begin{cases}
    \text{Interest} & \text{If no echo chambers}\\
    0 & \text{Else}
    \end{cases}
\end{equation}
It is required to obtain the interest of all possible user-item interactions, which cannot be satisfied by \textit{Movielens-1m} and \textit{Electronics}. Therefore, we use KuaiRec\footnote{https://chongminggao.github.io/KuaiRec/} dataest, which is a real-world dataset that contains a fully observed user-item interaction matrix (i.e., each user has feedback on each video). We define the interest of a user-item interaction as the video watching ratio which is the ratio of watching time length to the total video length. During the training, we consider watching ratio $\geq 2.0$ as positive feedback (like) and watching ratio $< 2.0$ as negative feedback (dislike) as noted by authors. Following the exit mechanism designed by \cite{gao2022cirs}, the interaction process ends if the user feels bored. Specifically, for the most recent $N$ recommended items, if at least $N_q$ items in the $N$ items share at least one attribute with the current recommended item, then the interaction process ends. Same as \cite{gao2022cirs}, we set $N$ as 1 and $N_q$ as 1, and apply softmax sampling to generate the recommendation, then add the recommended item to users' history for the next round recommendation. We report the average cumulative satisfactions and the interaction length in Table \ref{tab:satisfaction}. The MMR framework is not included because the framework is a re-ranking framework, which has the same prediction results as the original model when recommending one item in each round. 

\begin{table}[t]
    \centering
    \setlength{\tabcolsep}{8.pt}
    \begin{tabular}{c|cccccc}
    \toprule
        \multirow{2}{*}{} & \multicolumn{2}{c}{GRU4Rec} & \multicolumn{2}{c}{STAMP} & \multicolumn{2}{c}{NCR} \\\cmidrule(lr){2-3}\cmidrule(lr){4-5}\cmidrule(lr){6-7}
         & CS & Len & CS & Len & CS & Len \\\midrule
         Original & 1.333 & 26 & 2.987 & 46 & 2.519 & 49\\
         IPS & 2.159 & 40 & 1.827 & 35 & 1.412 & 34\\
         CausE & 1.857 & 32 & 2.734 & 42 & 2.459 & 47\\
         CCF & 2.577 & 43 & 3.152 & 38 & 2.534 & 49\\
         DICE & 1.994 & 36 & 2.859 & 44 & 2.830 & 52\\
         MACR & 2.960 & 38 & 3.036 & 52 & \textbf{3.107} & 37\\
         $\partial$CCF & \textbf{3.597} & \textbf{54} & \textbf{3.219} & \textbf{58} & 2.998 & \textbf{57} \\
         \bottomrule
    \end{tabular}
    \caption{The average cumulative satisfaction and interaction length on KuaiRec dataset. CS represents cumulative satisfaction and Len represents interaction length. The best performance is highlighted.}
    \vspace{-25pt}
    \label{tab:satisfaction}
\end{table}

From the results in Table \ref{tab:satisfaction}, we can observe that the $\partial$CCF framework achieves the best performance on cumulative satisfaction and interaction length in most cases. Without considering the echo chambers, the model may have higher satisfaction in a single round. However, such recommendation will narrow down to certain contents leading to echo chambers, which in turn makes users feel bored and quit. In this case, it may actually hurt users' satisfaction in the long run. Instead, the $\partial$CCF framework may help users to explore different content to mitigate echo chambers, which may not achieve the best satisfaction in a single round but will obtain higher user satisfaction in the long run.

\subsection{Sensitivity Analysis}\label{sec:sensitivity}
In this section, we will discuss the influence of the two important parameters in our framework from the perspective of recommendation performance and mitigating echo chambers. One is the number of counterfactual histories (i.e., $n$ in Eq.\eqref{eq:expectation}). The other is the value of factual probability (i.e., $\alpha$ in Eq.\eqref{eq:expectation}).

\subsubsection{\textbf{The number of counterfactual histories}}
Intuitively, in our model, each counterfactual history represents a possible result of the user actively exploring different items in the counterfactual world.
We change the number of counterfactual histories while keeping other parameters fixed. The results are in Figure \ref{fig:ctf_num}, including recommendation performance and the effect of echo chambers. 

For the recommendation performance (i.e., (a) and (c) in Figure \ref{fig:ctf_num}), we can observe that when the number of counterfactual histories is small (larger than 0), the recommendation performance is better than the base model in most cases, which means that proper counterfactual reasoning will make recommendation models better capture users' preference thus improve recommendation performance. However, when the number of counterfactual histories is too large, given that the factual probability is fixed, the probability of each counterfactual history (i.e., $\beta$ in Eq.\eqref{eq:uniform}) is small. In this case, 
too many counterfactual histories 
may introduce too much noise into the recommendation models and thus hurt the recommendation performance. For the performance on mitigating echo chambers (i.e., (b) and (d) in Figure \ref{fig:ctf_num}), we can observe that introducing counterfactual histories is helpful in mitigating echo chambers in most cases. However, when the number of counterfactual histories is too large, the items in counterfactual histories may not be well trained because the probability of each counterfactual history is too small, which will lead to worse performance on mitigating echo chambers. A proper number of counterfactual histories is required to mitigate the echo chambers without compromising the recommendation performance. 

\subsubsection{\textbf{The value of factual probability}}
The value of factual probability determines the distribution over factual and counterfactual histories. 
A larger value of factual probability represents more weights over the factual history. We tune the value of factual probability $\alpha$ while keeping the other parameters fixed. We plot the recommendation performance as well as the effect of echo chambers in Figure \ref{fig:fact_prob}. We can see that when the value is too small, the counterfactual histories will dominate the predictions, and thus may mislead the recommendation, but the framework still can mitigate the echo chambers in most cases. When the value is close to 1, the factual histories will dominate the predictions, thus the recommendation performance and the performance on mitigating echo chambers are closer to the base models.

\section{Conclusions and Future Work}\label{sec:conclusion}
In this paper, we show that DAG causal graphs cannot fully describe the dynamic data-generation process. As a result, we design a directed cyclic causal graph to represent the dynamic nature of recommender systems. Besides, the dynamic process may result in unwanted effects such as echo chambers. To theoretically understand echo chambers, we represent user-system interaction as a Markov Process and group users into three categories based on users' behaviors. Mathematically, for each group of users, we analyze whether the recommendation will lead to an echo chamber. In addition to the above theoretical contributions, we also design a Dynamic Causal Collaborative Filtering ($\partial$CCF) framework, which takes the back-door adjustment to estimate user preferences and leverages counterfactual reasoning to mitigate echo chambers. Experiments on real-world datasets show that our framework can mitigate echo chambers while achieving comparable recommendation performance with the base models. 

$\partial$CCF is a flexible framework for dynamic analysis of intelligent systems. In this work, we studied the echo chamber of recommender systems, while in the future, we can take $\partial$CCF to explore other dynamics such as popularity shifting, influence analysis and emerging topics. We can also conduct dynamic analysis of other systems such as social networks, online forums and conversational AI.

\subsubsection*{\textbf{Acknowledgment}}
This work was supported in part by NSF IIS 1910154, 2007907 and 2046457. Any opinions, findings, conclusions or recommendations expressed in this material are those of the authors and do not necessarily reflect those of the sponsors.

\bibliographystyle{ACM-Reference-Format}
\bibliography{reference}

\end{document}